# A Parametric Time Frequency-Conditional Granger Causality Method Using Ultra-regularized Orthogonal Least Squares and Multiwavelets for Dynamic Connectivity Analysis in EEGs

Yang Li, Mengying Lei*, Weigang Cui, Yuzhu Guo, and Hua-Liang Wei*

*Abstract*— ***Objective:*** **This study proposes a new parametric TF (time-frequency)-CGC (conditional Granger causality) method for high-precision connectivity analysis over time and frequency in multivariate coupling nonstationary systems, and applies it to scalp and source EEG signals to reveal dynamic interaction patterns in oscillatory neocortical sensorimotor networks.** ***Methods:*** **The Geweke's spectral measure is combined with the TVARX (time-varying autoregressive with exogenous input) modelling approach, which uses multiwavelets and ultra-regularized orthogonal least squares (UROLS) algorithm aided by APRESS (adjustable prediction error sum of squares), to obtain high-resolution time-varying CGC representations. The UROLS-APRESS algorithm, which adopts both the regularization technique and the ultra-least squares criterion to measure not only the signal data themselves but also the weak derivatives of them, is a novel powerful method in constructing time-varying models with good generalization performance, and can accurately track smooth and fast changing causalities. The generalized measurement based on CGC decomposition is able to eliminate indirect influences in multivariate systems.** ***Results:*** **The proposed method is validated on two simulations and then applied to multichannel motor imagery (MI)-EEG signals at scalp- and source-level, where the predicted distributions are well recovered with high TF precision, and the detected connectivity patterns of MI-EEG data are physiologically and anatomically interpretable and yield new insights into the dynamical organization of oscillatory cortical networks.** ***Conclusion:*** **Experimental results confirm the effectiveness of the proposed TF-CGC method in tracking rapidly varying causalities of EEG-based oscillatory networks.** ***Significance:*** **The novel TF-CGC method is expected to provide important information of neural mechanisms of perception and cognition.**

Manuscript received June 28, 2018. This work was supported by the National Natural Science Foundation of China [61671042, 61403016], Beijing Natural Science Foundation [4172037], Open Fund Project of Fujian Provincial Key Laboratory in Minjiang University [MJUKF201702], and Engineering and Physical Sciences Research Council (EPSRC) under Grant EP/I011056/1 and Platform programme under Grant EP/H00453X/1, U.K. (Corresponding authors: Mengying Lei; Hua-Liang Wei.)

Yang Li is with the Department of Automation Sciences and Electrical Engineering, Beijing Advanced Innovation Center for Big Data and Brain Computing, Beijing Advanced Innovation Center for Big Date-based Precision Medicine, Beihang University, Beijing, China.

Mengying Lei, Weigang Cui, and Yuzhu Guo are with the Department of Automation Sciences and Electrical Engineering, Beihang University, Beijing, China (e-mail: lmylei@buaa.edu.cn).

Hua-Liang Wei is with the Department of Automatic Control and Systems Engineering, The University of Sheffield, Sheffield S1 3DJ, U.K (e-mail: w.hualiang@sheffield.ac.uk).

## I. INTRODUCTION

DYNAMIC interactions within brain regions enable synchronization of neuronal oscillations, which is a suggested mechanism underlying the perceptual and cognitive functions [1]. Analyzing time-varying interaction patterns of oscillatory brain networks is a considerably important and challenging research topic in the neuroscience field [2]. Recently, dynamic Granger causality (GC) [3] analysis has emerged as a powerful technique to detect directed interactions among coupled nonstationary systems, and has been extensively investigated in neurophysiological studies [4, 5]. The key in dynamic GC detection is the identification of the time-varying autoregressive with exogenous input (TVARX) models for nonstationary signals. Several methods have been developed for assessing dynamic GC relations in time or frequency domain [6], mainly including nonparametric method [7, 8], sliding window approach [9], adaptive multivariate estimation [10] and parametric modelling approach [11-13].

In the nonparametric GC detection method proposed in [7], the time-frequency (TF) causality analysis was based on nonparametric wavelet transforms and the performance demonstrated on monkey local field potentials. Nevertheless, for this method, it is difficult to select desirable initial parameters to simultaneously ensure both good time and frequency resolution, and thus the estimates may not be reliable when only a few trials data sets of short length are available. In the sliding window approach (e.g. [14]), the temporal functions of spectral GC can be roughly extracted by analyzing traditional time-invariant GC influences for each single window through ARX modelling. However, the time resolution of this approach is smeared and the detection performance depends on the window size; this limits its practical applicability. In the adaptive multivariate strategy, the recursive least squares (RLS) and Kalman filtering algorithms are the most commonly used approaches for the estimation of time-varying parameters [10, 15]. These adaptive methods can detect slow varying interaction relations in TF domain, but they are sensitive to noises and may fail to track rapid changing connectivity due to the slow convergence speed.

Compared with the above mentioned methods, the parametric approach, based upon TVARX model identification using a basis function expansion scheme, can usually provide better performance for dynamic GC detection. In such a detection framework, the basic time-varying models of signals are firstly estimated by applying a set of pre-defined basis functions with good representation properties [16, 17] and running an efficient model structure determination algorithm such as the orthogonal forward regression [18, 19]; time-varying variances of model prediction errors and corresponding GCs can then be effectively calculated from the reduced refined TVARX models. For example, Li *et al.* employed multiwavelet basis functions with regularized orthogonal least squares (ROLS) to approximate the time-varying parameters of TVARX models, which were applied in successfully detecting both rapid and slow varying causalities between two nonstationary signals [20].

Despite the multiwavelet expansion approach with ROLS algorithm provides a general parametric method for time-varying GC detection, two deficiencies are remained in this scheme. First, although the ROLS algorithm enables better generalization in model construction than the classic OLS and works well even in the presence of severe noises [21, 22], the method may produce suboptimal model with possible spurious or insufficient model terms when the signals are not persistently exciting or contaminated by different levels of noises [23, 24]. In this case, the resulting under-fitting TVARX models might produce incorrect and low precision GC distributions. Second, this pairwise time-domain GC approach ignores frequency information which is crucial for the analysis of neurophysiological signals with abundant oscillatory content, like electroencephalography (EEG), and it cannot distinguish direct and indirect effects among systems with more than two simultaneously acquired signals. Thus, the conventional ROLS method may fail to reveal dynamic connectivity in coupled oscillatory brain networks. Currently, EEG technique is often used for studying brain activities, because of its non-invasive nature, good temporal resolution and low cost [25]. However, there is still lack of high resolution time-frequency causality method for EEG-based connectivity analysis even in recent researches due to the high nonstationarity and complexity of EEG signals.

In this paper, we propose a new parametric TF-CGC (time-frequency conditional Granger causality) method for analyzing dynamic connectivity among multivariate coupling nonstationary systems over time and frequency, where the TVARX modeling approach, implemented by a powerful ultra-regularized orthogonal least squares (UROLS) algorithm, is combined with the spectral CGC measure to obtain the time-frequency causality analysis. The time-varying parameters in TVARX models are firstly expanded by a finite number of multiwavelet basis functions for tracking both the overall global trend and transient local changes in nonstationary signals [24, 26]. Then the UROLS algorithm, which improves the conventional ROLS in using not only the residuals between the observed signals and the predicted values but also the associated weak derivatives to measure the model fitness [23], is applied to determine the parsimonious model structure and associated parameters. In the proposed UROLS algorithm, a modified cross-validation criterion named adjustable prediction error sum of squares (APRESS) is incorporated to facilitate the monitoring of the forward orthogonal search procedure and the determination of the model complexity [27, 34]. Finally, a high resolution TF-CGC representation is established by combining the accurately identified TVARX models with the statistically-explicable mathematical framework of Geweke's spectral CGC [28]. Our proposed TF-CGC method is firstly tested on two simulated nonstationary coupling systems, and then it is applied to scalp EEG data acquired from MI tasks and the corresponding source signals. Experimental results demonstrate the efficiency of the proposed TF-CGC method for detecting dynamic interaction activities among nonstationary and oscillatory brain systems. A main contribution of the study is that for the first time the parametric TVARX modeling approach is introduced and combined with spectral CGC decomposition to obtain a high-resolution representation of dynamic CGC in TF domain. It is worth mentioning that the newly proposed UROLS-APRESS algorithm is innovatively applied with multiwavelet-based modelling scheme for effectively identifying TVARX models. It is expected that the novel implementation of the UROLS with multiwavelets to TF-CGC analysis can provide important insights into the neural mechanisms underlying perceptual and cognitive functions, and inspire further development of more powerful approaches for dynamic connectivity analysis.

## II. METHODS

The concept of classic GC is formulated based on univariate AR or bivariate ARX models. A TF-CGC decomposition method which combines the time-varying system identification approach with Geweke's spectral CGC measure, is proposed in this work. The TF-CGC decomposition for multivariate time series is built on TVARX modelling, thus the newly introduced nonstationary model identification method is firstly discussed in this section. The discussion focus on a case involving three time series, but it can easily be extended to more than three sets of time series.

*A. TVARX model identification using multiwavelets for TF-CGC analysis*

Consider three stochastic processes $X = \{x(t)\}, Y = \{y(t)\}$ and $Z = \{z(t)\}$, with sampling index $t = 1, 2, \cdots, N$, where the TF-CGC relations from $Y$ to $X$ conditional on $Z$ is to be evaluated. Let the joint TVARX representations of $x(t)$ and $z(t)$ be

$$x(t) = \sum_{i=1}^{l_1} a_{11,i}(t) x(t-i) + \sum_{i=1}^{l_2} a_{12,i}(t) z(t-i) + e_1(t)$$
$$z(t) = \sum_{i=1}^{l_1} a_{21,i}(t) x(t-i) + \sum_{i=1}^{l_2} a_{22,i}(t) z(t-i) + e_2(t)$$
(1)

Denote the joint TVARX model of all the three processes $x(t)$, $y(t)$ and $z(t)$ as

$$x(t) = \sum_{i=1}^{l_1} b_{11,i}(t) x(t-i) + \sum_{i=1}^{l_2} b_{12,i}(t) y(t-i) + \sum_{i=1}^{l_3} b_{13,i}(t) z(t-i) + e_3(t)$$
$$y(t) = \sum_{i=1}^{l_1} b_{21,i}(t) x(t-i) + \sum_{i=1}^{l_2} b_{22,i}(t) y(t-i) + \sum_{i=1}^{l_3} b_{23,i}(t) z(t-i) + e_4(t)$$
$$z(t) = \sum_{i=1}^{l_1} b_{31,i}(t) x(t-i) + \sum_{i=1}^{l_2} b_{32,i}(t) y(t-i) + \sum_{i=1}^{l_3} b_{33,i}(t) z(t-i) + e_5(t)$$
(2)

An efficient solution when identifying these TVARX models is to expand the time-varying parameters onto a set of basis functions $\{\varphi_m(t): m = 1,2,\cdots,M\}$, specifically for the trivariate TVARX process with respect to signal $x(t)$ defined in (2)

$$x(t) = \sum_{i=1}^{I_1} b_{11,i}(t)x(t-i) + \sum_{i=1}^{I_2} b_{12,i}(t)y(t-i)$$
$$+ \sum_{i=1}^{I_3} b_{13,i}(t)z(t-i) + e_3(t) \quad (3)$$
$$= \sum_{n=1}^{V} \sum_{i=1}^{I} c_{n,i}(t) \prod_{n=1} x(t-i) \prod_{n=2} y(t-i) \prod_{n=3} z(t-i) + e_3(t)$$

where $x(t)$, $y(t)$, $z(t)$ are the system output and input with maximum lags $I_1$, $I_2$ and $I_3$, respectively, $V = 3$ is the number of input variables, $\{c_{1,i}(t)\}_{i=1}^{I}$, $\{c_{2,i}(t)\}_{i=1}^{I}$ and $\{c_{3,i}(t)\}_{i=1}^{I}$ are the time-varying parameters to be determined, $e_3(t)$ is assumed to be a sequence of independent and normal distributed random variables with zero mean; then this TVARX model can be expanded as

$$x(t) = \sum_{n=1}^{V}\sum_{i=1}^{I}\sum_{m=1}^{M} \alpha_{n,i,m}\varphi_m(t) \prod_{n=1} x(t-i) \prod_{n=2} y(t-i) \prod_{n=3} z(t-i) + e_3(t)$$
$$= \psi(t)^T \theta + e_3(t) \quad (4)$$

where $\alpha_{n,i,m}$ denote the time-invariant expansion parameters of basis functions $\varphi_m(t)$, $M$ is the number of the basis sequences, $\psi(t) = [\chi_X(t), \chi_Y(t), \chi_Z(t)]^T$ is a $(I_1 + I_2 + I_3) \times M \times 1$ dimensional regression vector, in which $\chi_X(t) = [x(t-1)\phi(t)^T, x(t-2)\phi(t)^T, \cdots, x(t-I_1)\phi(t)^T]$, $\chi_Y(t) = [y(t-1)\phi(t)^T, y(t-2)\phi(t)^T, \cdots, y(t-I_2)\phi(t)^T]$, and $\chi_Z(t) = [z(t-1)\phi(t)^T, z(t-2)\phi(t)^T, \cdots, z(t-I_3)\phi(t)^T]$ with $\phi(t) = [\varphi_1(t), \varphi_2(t), \cdots, \varphi_M(t)]^T$, the expansion coefficient vector is $\theta = [\alpha_{1,1,1}, \cdots, \alpha_{1,I,M}, \cdots, \alpha_{V,1,1}, \cdots \alpha_{V,I,M}]^T$, and the upper script $T$ represents the transpose of a vector. The initial time-varying model then becomes a time-invariant regression problem, since all $\alpha_{n,i,m}$ are now time invariant.

In practice, a proper selection of the basis functions is vital to ensure the identified model performance. A good suggestion is to use multiple wavelet basis functions to effectively track both rapid and slow parameter variations in time-varying processes [19]. This work suggests using multi-wavelet basis functions to approximate the time-varying parameters in (3) as

$$c_{n,i}(t) = \sum_r \sum_{k \in \Gamma_r} \beta_{n,i,k}^r \xi_{k,j}^r\left(\frac{t}{N}\right) \quad (5)$$

where $\xi_{k,j}^r(\cdot)$ are wavelet basis functions, with the shift indices $k \in \Gamma_r, \Gamma_r = \{k: -r \le k \le 2^j - 1\}$ and wavelet scale $j$, $\beta_{n,i,k}^r$ represent the corresponding expanded basis function parameters which are time invariant, $r$ denotes the order of the wavelet basis functions, and the function variable $t/N$ is normalised within $[0,1]$.

Cardinal B-splines are an important class of basis functions that simultaneously possess three remarkable properties, namely compactly supported, analytically formulated and multiresolution analysis oriented, which enable the operation of the wavelet decomposition to be more convenient [29]. Taking the cardinal B-splines as the basis function, the $\xi_{k,j}^r(\cdot)$ can be expressed by the $r$-th order B-spline $B_r$ as $\xi_{k,j}^r(u) = 2^{j/2} B_r(2^j u - k)$, where $j, k$ are the dilated and shifted versions of wavelet $B_r$. Generally $j$ is chose to be 3 or a larger number in many B-splines applications [26], and a practical selection of the wavelets are $\{\xi_{k,j}^r: r = 3,4,5\}$, the detail description of B-splines properties can be found in [30]. The decomposition (5) can easily be transformed into the form of (4), where the collection $\{\varphi_m(t): m = 1,2,\cdots,M\}$ is replaced by the union of multi-B-splines families $\sum_r \sum_{k \in \Gamma_r} \xi_{k,j}^r(u)$, then the TVARX model becomes

$$x(t) = \sum_{n=1}^{V}\sum_{i=1}^{I}\sum_r \sum_{k \in \Gamma_r} \beta_{n,i,k}^r \xi_{k,j}^r\left(\frac{t}{N}\right)$$
$$\times \prod_{n=1} x(t-i) \prod_{n=2} y(t-i) \prod_{n=3} z(t-i) + e_3(t) \quad (6)$$
$$= \Psi^T(t)\delta + e_3(t)$$

where $\Psi^T(t)$ is the expanded term vector at time $t$ and $\delta$ is the corresponding time-invariant parameter vector.

Equation (6) indicates that the multi-wavelet basis function expansion method converts the identification of the time-varying model (3) to solving a time-invariant regression problem. However, the number of candidate model terms in $\Psi^T(t)$ can be very large if the number of involved wavelet basis functions $r$, the wavelet scale $j$ or the maximum lags $I_1, I_2, I_3$ are large; as a consequence, the initial full regression model (6) is often redundant, ill-conditioned and not ready for direct use. Thus, selecting significant terms from the pool of the expanded regressors and building a sparse model structure is highly required, and this will be introduced in the next section.

### B. The UROLS algorithm for TVARX model identification in TF-CGC analysis

The identification of the TVARX model includes two steps: determining the model structure and estimating the associated parameters. In this section, a new method, referred to as ultra-regularized orthogonal least squares (UROLS), is proposed for time-varying model identification; it incorporates the following three approaches: the ultra least squares (ULS) metric, the regularized orthogonal least squares (ROLS) algorithm, and adjustable prediction error sum of squares (APRESS).

For generic regression problem, the least squares loss function aims to achieve best model fitting on the Lebesgue space $L^2([0,T])$, where $[0,T]$ is the time span of the signals, and the model that minimizes the square of the $L^2$ norm is to be identified. The $L^2$ norm, only measures the similarity of two functions as a whole, cannot characterize the local distribution difference at each time instance, thus neglects some important information of details in shape [23]. The absence of this crucial information might lead to a model structure which cannot sufficiently represent the inherent dynamics of the data (and therefore the associated system) especially when the system is not persistently excited. It is known that most physical systems behave mainly as a low-pass filter, and are actually defined on the subspace of $L^2([0,T])$, that is, the Sobolev space $H^d([0,T]) = W^{d,2}([0,T])$, $H^d([0,T]) = \{u(t) \in L^2([0,T]) | D^v u \in L^2([0,T]), v = 1,2,\cdots,d\}$, where the weak derivatives $D^v u$ up to $d$-th are also $L^2$ integrable [17]. Thus, a stricter metric, which can reveal the entire useful information of observations realized in the Sobolev space, is used in this study. Note that much of such information is ignored in nearly all existing model structure detection algorithms based on traditional least squares.

A stricter metric for $H^d([0,T])$ is the $H^d$ norm defined as $u_{H^d} = \sqrt{\sum_{v=1}^d \|D^v u\|_2^2}$ [23]. Based on this norm, a new criterion for model (6) can be defined as

$$J_H = \left\| x - \sum_{n=1}^{V}\sum_{i=1}^{1}\sum_{r}\sum_{k \in \Gamma_r} \beta_{n,i,k}^r u_{n,i,k}^r \right\|_2^2 +$$

$$\sum_{\upsilon=1}^{d} \left\| \left\{ D^\upsilon x - \sum_{n=1}^{V}\sum_{i=1}^{1}\sum_{r}\sum_{k \in \Gamma_r} \beta_{n,i,k}^r D^\upsilon u_{n,i,k}^r \right\} \right\|_2^2 \quad (7)$$

$$= J_{H_1} + J_{H_2}$$

where $u_{n,i,k}^r(t) = \xi_{k,j}^r(t/N) \prod_{n=1} x(t-i) \prod_{n=2} y(t-i) \times \prod_{n=3} z(t-i)$ are the expanded terms. This loss function contains two parts: the first part is the standard least squares criterion which focuses on the similarity over the whole data, while the second part describes the identity of the weak derivatives which essentially emphases the unity in shape. The second part, which fully takes into account the agreement in shape of signals, makes this new criterion different to most traditional methods for model structure detection. Any detailed difference in the distribution can be characterized in the second part of the new cost function (7). Thus, the criterion $J_H$ is a more effective metric for model identification than the conventional least squares criterion. According to the $H^d$ norm, the regression problem can be converted to solve a new ULS problem

$$\begin{bmatrix} x \\ D^1 x \\ \vdots \\ D^d x \end{bmatrix} = \sum_{n=1}^{V}\sum_{i=1}^{1}\sum_{r}\sum_{k \in \Gamma_r} \beta_{n,i,k}^r \begin{bmatrix} u_{n,i,k}^r \\ D^1 u_{n,i,k}^r \\ \vdots \\ D^d u_{n,i,k}^r \end{bmatrix} \quad (8)$$

However, note that in many practical systems the weak derivatives are not able to be directly calculated from the observed data, and the contribution of each component in (7) might be quite different because the amplitude of the derivatives could be much larger than that of the errors resulting from the data themselves, i.e. $J_{H_2} \gg J_{H_1}$, especially when the residuals change rapidly. A consequence would be that the effect of noise is magnified improperly dominate the whole metric (7).

In order to properly assess the contribution of the unknown weak derivatives in $J_H$, the distributions of $x$ and $u_{n,i,k}^r$ are introduced and defined as

$$x^\upsilon(\tau) \triangleq (-1)^\upsilon \int_0^T x(t) \omega^{(\upsilon)}(t-\tau) dt$$
$$\left( u_{n,i,k}^r \right)^\upsilon(\tau) \triangleq (-1)^\upsilon \int_0^T u_{n,i,k}^r(t) \omega^{(\upsilon)}(t-\tau) dt \quad (9)$$

where $\omega(t)$ is a test function with a finite support on $[0, T_0]$, $T_0 < T$ and time shift $\tau \in [0, T - T_0]$ [23]. Here $x^\upsilon(\tau)$ is the convolution of $x(t)$ with the $\upsilon$-th derivative of the test function. Set $g(t) = \omega(-t)$, then $g^{(\upsilon)}(t)$ can be regarded as the impulse response of a linear filter and $x^\upsilon(\tau)$ is the filter output of $x(t)$. The function $x^\upsilon(\tau)$ now gets a new physical interpretation which represents a signal obtained by smoothing first and then evaluating the derivatives of the smoothed signals.

Additionally, the test function and corresponding derivatives are further normalized as $\bar{\omega}^{(\upsilon)} = \omega^{(\upsilon)}/\|\omega^{(\upsilon)}\|_2, \upsilon = 1,2,\cdots,d$ to prevent the derivative part from dominating $J_H$ in (7) and make this criterion robust to noise. Appling these normalized test functions to modulate signals in the ULS problem, it can ensure that each data from the modulated function $x^\upsilon(\tau)$ has the same weight as the data in $x(t)$. Besides, the test function should have a bell shape like the Gaussian function for smoothing the initial signals. This study uses the $(d+1)$-th order B-splines which have finite support and continuous $d$-th order derivatives as the test functions. Given a test function $\omega(t)$, the ULS criterion can then be expressed as

$$J_{ULS} = \left\| x - \sum_{n=1}^{V}\sum_{i=1}^{1}\sum_{r}\sum_{k \in \Gamma_r} \beta_{n,i,k}^r u_{n,i,k}^r \right\|_2^2$$
$$+ \sum_{\upsilon=1}^{d} \left\| \left\{ \bar{x}^\upsilon - \sum_{n=1}^{V}\sum_{i=1}^{1}\sum_{r}\sum_{k \in \Gamma_r} \beta_{n,i,k}^r \left( \bar{u}_{n,i,k}^r \right)^\upsilon \right\} \right\|_2^2 \quad (10)$$

where $\bar{x}^\upsilon(\tau) = \int_0^t x(t) \bar{\omega}^{(\upsilon)}(t-\tau) dt$ and $\left( \bar{u}_{n,i,k}^r \right)^\upsilon(\tau) = \int_0^t u_{n,i,k}^r(t) \bar{\omega}^{(\upsilon)}(t-\tau) dt$. Given sampled data with discrete time $t = 1,2,\cdots,N$, the discrete form of the modulating procedure is denoted as $\bar{x}^\upsilon(p) = \sum_{t=p}^{p+n_0} x(t) \bar{\omega}^\upsilon(t-p)$ and $\left( \bar{u}_{n,i,k}^r \right)^\upsilon(p) = \sum_{t=p}^{p+n_0} u_{n,i,k}^r(t) \bar{\omega}^{(\upsilon)}(t-p)$, where $n_0$ is the support of the test function and $p = 1,2,\cdots,N-n_0$. Then the matrix form of the ULS problem can be represented as

$$X_{ULS} = \Phi_{ULS} \Theta \quad (11)$$

where

$$X_{ULS} = \left[ x(1),\cdots,x(N), \bar{x}^1(1),\cdots,\bar{x}^d(N-n_0) \right]^T \quad (12)$$

$$\Phi_{ULS} = \begin{bmatrix} u_{1,i,k}^r(1) & \cdots & u_{V,i,k}^r(1) \\ \vdots & \cdots & \vdots \\ u_{1,i,k}^r(N) & \cdots & u_{V,i,k}^r(N) \\ \left( \bar{u}_{1,i,k}^r \right)^1(1) & \cdots & \left( \bar{u}_{V,i,k}^r \right)^1(1) \\ \vdots & \cdots & \vdots \\ \left( \bar{u}_{1,i,k}^r \right)^d(N) & \cdots & \left( \bar{u}_{V,i,k}^r \right)^d(N) \end{bmatrix} \quad (13)$$

$$\Theta = \left[ \beta_{1,i,k}^r,\cdots,\beta_{V,i,k}^r \right]^T \quad (14)$$

Now the TVARX model (3) is transformed into another problem of constructing model (11), which can be solved by means of a model structure detection method such as the well-known forward regression OLS algorithm [31, 32]. Although OLS has proven to be an efficient procedure for model construction and refinement, the use of the parsimonious principle alone cannot entirely avoid overfitting since small-sized models constructed may still fit to the noise when the systems are highly noisy [33]. In order to alleviate such a dilemma, an effective zero-order ROLS ($ROLS^0$) technique combining the zero-order regularization with the OLS [21, 22] is used in this study, with which a sparse model structure with good generalization performances and low computational costs can be constructed.

For the regression model (11), $X_{ULS}$ is a vector of system outputs and $\Phi_{ULS}$ is a matrix formed by candidate terms (regressors). Denote all the candidate bases by a dictionary $D = \{\gamma_{n,i,m}: n = 1,\cdots V; i = 1,\cdots I; m = 1,\cdots M\}$, where $\gamma_{n,i,m}(t) = u_{n,i,k}^r(t)$, and the term selection procedure is to find a full dimensional subset $D_\eta = \{\gamma_{L_\kappa}: \kappa = 1,2,\cdots,\eta; L_\kappa \in \{1,2,\cdots (I_1 + I_2 + I_3) \times M\}\} (\eta \ll (I_1 + I_2 + I_3) \times M)$, so that $X$ can be approximated via a linear combination of $\gamma_{L_\kappa}$ as $X = \gamma_{L_1} \pi_{L_1} + \cdots + \gamma_{L_\eta} \pi_{L_\eta} + e$ or in a compact matrix form $X = \Upsilon \Pi + e$, where the regression matrix $\Upsilon = [\gamma_{L_1}, \gamma_{L_2}, \cdots, \gamma_{L_\eta}]$ is of full column rank, $\Pi = [\pi_{L_1}, \pi_{L_2}, \cdots, \pi_{L_\eta}]^T$ is the associated parameter vector and $e = [e_3(1), e_3(2), \cdots, e_3(N)]^T$ is the approximation error vector.

For constructing a parsimonious model structure, the $ROLS^0$ algorithm is used to orthogonalize the candidate model terms and determine significant terms according to the zero-order regularized error reduction ratio ($RERR^0$) defined as [21]

$$RERR^0(X,\gamma) = \frac{\langle X,\gamma \rangle^2}{\langle X,X \rangle (\langle \gamma,\gamma \rangle + \mu)} \quad (15)$$

where $X$ and $\gamma$ are the output vector and a candidate term respectively, $\mu \geq 0$ is the regularization parameter, and the symbol $\langle \cdot, \cdot \rangle$ denotes the inner product of two vectors. The first significant term can be determined by $L_1 = arg \max_{1 \leq l \leq (I_1+I_2+I_3) \times M} \{RERR^0(X,\gamma_1)\}$, and the first corresponding orthogonal basis can be selected as $h_1 = \gamma_{L_1}$. Assume that a subset $D_{\varsigma-1}$ consisting of $(\varsigma - 1)$ significant terms $\gamma_{L_1}, \gamma_{L_2}, \cdots, \gamma_{L_{\varsigma-1}}$ have been chosen at step $(\varsigma - 1)$, and these terms can be transformed into a group of orthogonalised bases $h_1, h_2, \cdots, h_{\varsigma-1}$, generally the $\varsigma$-th significant term can be chosen as follows. Set

$$h_q^{(\varsigma)} = \gamma_q - \sum_{s=1}^{\varsigma-1} \frac{\gamma_q^T h_s}{h_s^T h_s} h_s \quad (16)$$

where $\gamma_q \in D - D_{\varsigma-1}$, and the $\varsigma$-th significant term can be selected by $L_\varsigma = arg \max_{q \in l, q \neq L_s, 1 \leq s \leq \varsigma-1} \{RERR^0(X, h_q^{(\varsigma)})\}$. Then the $\varsigma$-th significant basis is chosen as $\gamma_{L_\varsigma}$ and the associated orthogonal basis is $h_\varsigma = h_{L_\varsigma}^{(\varsigma)}$. Subsequent significant terms can be selected similarly in a forward regression manner based on the $RERR^0$. Additionally, a modified cross-validation criterion named adjustable prediction error sum of squares (APRESS) defined bellow is integrated into the UROLS algorithm to decide the termination of the term search procedure [27, 34]

$$APRESS(g) = p(g)\left[\|r_g\|^2 / N\right] \quad (17)$$

where $p(g) = 1/(1 - g\nu/N)^2$ with adjustable parameter $\nu \geq 1$ is the penalty function, $\|r_g\|^2 = \|X\|^2 - \sum_{\varsigma=1}^{g} \frac{(r_{\varsigma-1}^T h_\varsigma)^2}{h_\varsigma^T h_\varsigma}$, $r_0 = X$ is the residual sum of squares, and $\|r_g\|^2/N$ denotes the mean-squared-errors (MSE) obtained from the associated $g$-term model. The term selection process is terminated when the APRESS statistic arrives at minimum. The effect of $\nu$ on results is detailed in [34]. The selected regression matrix $\Upsilon = [\gamma_{L_1}, \gamma_{L_2}, \cdots, \gamma_{L_\eta}]$ can be orthogonally decomposed as $\Upsilon = O_\eta R_\eta$, where $O_\eta$ is a $N \times \eta$ matrix with orthogonal columns and $R_\eta$ is a $\eta \times \eta$ unit upper triangular matrix. Then the corresponding parameter vector $\Pi = [\pi_{L_1}, \pi_{L_2}, \cdots, \pi_{L_\eta}]^T$ can be calculated from the formula $R_\eta \Pi = U$, where $U = (O_\eta^T O_\eta)^{-1} R_\eta^T X$, and the time-varying coefficients in the TVARX model (3) can thus be recovered using the resultant estimates. Similar to the TVARX model (3), other multivariate TVARX processes expressed in (1)-(2) can also be identified using the proposed multiwavelets and UROLS-APRESS method.

*C. The formulation of TF-CGC analysis*

The proposed UROLS method can provide more accurate TVARX models for nonstationary time series with respect to $x(t)$, $y(t)$ and $z(t)$ given in (1)-(2), and this is the most considerable basis of TF-CGC analysis. The formulation of TF-CGC from $Y$ to $X$ conditional on $Z$ denoted as $GC_{Y \to X|Z}(t,f)$ is provided in this section.

In (1), the initial noise terms $e_1(t)$ and $e_2(t)$ could be correlated with each other and their time-varying covariance matrix is $\Sigma_1 = [(\Sigma_1(t)\ \Delta_1(t)), (\Delta_1(t)\ \Sigma_2(t))]^T$, specifically $\Sigma_1(t) = var(e_1(t))$, $\Sigma_2(t) = var(e_2(t))$ and $\Delta_1(t) = cov(e_1(t), e_2(t))$ are calculated using a general recursive expression [11] $\sigma^2(t+1) = (1-\rho)\sigma^2(t) + \rho u_1(t)u_2(t)$ with $0 < \rho < 1$. Setting $u_1(t) = u_2(t) = e_1(t)$, $u_1(t) = u_2(t) = e_2(t)$, and $u_1(t) = e_1(t), u_2(t) = e_2(t)$, yields time-varying variances and covariance of the corresponding prediction errors $\Sigma_1(t)$, $\Sigma_2(t)$ and $\Delta_1(t)$, respectively. Define the lag operator $\lambda$ to be $\lambda x(t) = x(t - \lambda)$, then (1) can be rewritten as

$$\begin{pmatrix} a_{11}(\lambda) & a_{12}(\lambda) \\ a_{21}(\lambda) & a_{22}(\lambda) \end{pmatrix} \begin{pmatrix} x(t) \\ z(t) \end{pmatrix} = \begin{pmatrix} e_1(t) \\ e_2(t) \end{pmatrix} \quad (18)$$

where $a_{11}(0) = a_{22}(0) = 1$, $a_{12}(0) = a_{21}(0) = 0$. The independence of $e_1(t)$ and $e_2(t)$ is necessary for the definition of spectral domain causality [3]. Thus the normalization procedure introduced by Geweke [28] is exploited and developed to remove the correlation and further make the identification of an intrinsic part and a causal part possible in time-varying cases. The transformation consists of left-multiplying $P(t) = [(1\ 0), (-\Delta_1(t)/\Sigma_1(t)\ 1)]^T$ on both sides of (18) at each time index [35], and the resulting normalized form is given as

$$\begin{pmatrix} A_{11}(\lambda) & A_{12}(\lambda) \\ A_{21}(\lambda) & A_{22}(\lambda) \end{pmatrix} \begin{pmatrix} x(t) \\ z(t) \end{pmatrix} = \begin{pmatrix} \varepsilon_1(t) \\ \varepsilon_2(t) \end{pmatrix} \quad (19)$$

where $A_{11}(0) = A_{22}(0) = 1$, $A_{12}(0) = 0$, $A_{21}(0)$ is generally not zero, $cov(\varepsilon_1(t), \varepsilon_2(t)) = 0$, and note that $var(\varepsilon_1(t)) = \Sigma_1(t)$.

In (2), the time-varying covariance matrix of the noise terms can be estimated by the recursive computation similarly as $\Sigma_1$, and is given by $\Sigma_2 = [(\Sigma_{xx}(t), \Sigma_{xy}(t), \Sigma_{xz}(t)), (\Sigma_{yx}(t), \Sigma_{yy}(t), \Sigma_{yz}(t)), (\Sigma_{zx}(t), \Sigma_{zy}(t), \Sigma_{zz}(t))]^T$. The normalization process of (2) involves left-multiplying both sides by the time-varying matrix $Q(t) = Q_2(t) \cdot Q_1(t)$ at each discrete time [35], where

$$Q_1(t) = \begin{pmatrix} 1 & 0 & 0 \\ -\Sigma_{yx}(t)\Sigma_{xx}^{-1}(t) & 1 & 0 \\ -\Sigma_{zx}(t)\Sigma_{xx}^{-1}(t) & 0 & 1 \end{pmatrix} \quad (20)$$

$$Q_2(t) = \begin{pmatrix} 1 & 0 & 0 \\ 0 & 1 & 0 \\ 0 & -(\Sigma_{zy}(t)-\Sigma_{zx}(t)\Sigma_{xx}^{-1}(t)\Sigma_{xy}(t))(\Sigma_{yy}(t)-\Sigma_{yx}(t)\Sigma_{xx}^{-1}(t)\Sigma_{xy}(t))^{-1} & 1 \end{pmatrix} \quad (21)$$

Then the associated normalized equations for (2) can be expressed as

$$\begin{pmatrix} B_{11}(\lambda) & B_{12}(\lambda) & B_{13}(\lambda) \\ B_{21}(\lambda) & B_{22}(\lambda) & B_{23}(\lambda) \\ B_{31}(\lambda) & B_{32}(\lambda) & B_{33}(\lambda) \end{pmatrix} \begin{pmatrix} x(t) \\ y(t) \\ z(t) \end{pmatrix} = \begin{pmatrix} \varepsilon_3(t) \\ \varepsilon_4(t) \\ \varepsilon_5(t) \end{pmatrix} \quad (22)$$

where the noise terms are now independent to each other, and their time-varying variances are $\tilde{\Sigma}_{xx}(t)$, $\tilde{\Sigma}_{yy}(t)$ and $\tilde{\Sigma}_{zz}(t)$, respectively. According to the crucial relationship of conditional causality in time and frequency domain [28], the problem of measuring the time-dependent spectral causal connectivity $GC_{Y\to X|Z}(t,f)$ can thus be converted into the calculation of the causal influence from $Y\varepsilon_2$ to $\varepsilon_1$. In order to obtain $GC_{Y\varepsilon_2\to\varepsilon_1}(t,f)$, the variance of $\varepsilon_1$ is next decomposed in the time and frequency domain. Time-frequency transforming both sides of (19) leads to

$$\underbrace{\begin{pmatrix} A_{11}(t,f) & A_{12}(t,f) \\ A_{21}(t,f) & A_{22}(t,f) \end{pmatrix}}_{\boldsymbol{A}(t,f)} \begin{pmatrix} X(t,f) \\ Z(t,f) \end{pmatrix} = \begin{pmatrix} E_1(t,f) \\ E_2(t,f) \end{pmatrix} \quad (23)$$

where the components of the coefficient matrix $\boldsymbol{A}(t,f)$ are

$$A_{11}(t,f) = 1 - \sum_{i=1}^{l_1} A_{11,i} e^{-j_0 2\pi i f/f_s}, \quad A_{12}(t,f) = -\sum_{i=1}^{l_2} A_{12,i} e^{-j_0 2\pi i f/f_s}$$

$$A_{21}(t,f) = -\sum_{i=1}^{l_1} A_{21,i} e^{-j_0 2\pi i f/f_s}, \quad A_{22}(t,f) = 1 - \sum_{i=1}^{l_2} A_{22,i} e^{-j_0 2\pi i f/f_s}$$

with $j_0 = \sqrt{-1}$ and $f_s$ being the sampling frequency. Similarly, calculating the time-varying spectral decomposition of (22) and representing it as

$$\underbrace{\begin{pmatrix} B_{11}(t,f) & B_{12}(t,f) & B_{13}(t,f) \\ B_{21}(t,f) & B_{22}(t,f) & B_{23}(t,f) \\ B_{31}(t,f) & B_{32}(t,f) & B_{33}(t,f) \end{pmatrix}}_{\boldsymbol{B}(t,f)} \begin{pmatrix} X(t,f) \\ Y(t,f) \\ Z(t,f) \end{pmatrix} = \begin{pmatrix} E_3(t,f) \\ E_4(t,f) \\ E_5(t,f) \end{pmatrix} \quad (24)$$

Recasting (23) and (24) into the transfer function format we obtain

$$\begin{pmatrix} X(t,f) \\ Z(t,f) \end{pmatrix} = \underbrace{\begin{pmatrix} G_{xx}(t,f) & G_{xz}(t,f) \\ G_{zx}(t,f) & G_{zz}(t,f) \end{pmatrix}}_{\boldsymbol{G}(t,f)} \begin{pmatrix} E_1(t,f) \\ E_2(t,f) \end{pmatrix} \quad (25)$$

$$\begin{pmatrix} X(t,f) \\ Y(t,f) \\ Z(t,f) \end{pmatrix} = \underbrace{\begin{pmatrix} K_{xx}(t,f) & K_{xy}(t,f) & K_{xz}(t,f) \\ K_{yx}(t,f) & K_{yy}(t,f) & K_{yz}(t,f) \\ K_{zx}(t,f) & K_{zy}(t,f) & K_{zz}(t,f) \end{pmatrix}}_{\boldsymbol{K}(t,f)} \begin{pmatrix} E_3(t,f) \\ E_4(t,f) \\ E_5(t,f) \end{pmatrix} \quad (26)$$

where the TF transfer function $\boldsymbol{G}(t,f)$ and $\boldsymbol{K}(t,f)$ are the inverse of the normalized coefficient matrix $\boldsymbol{A}(t,f)$ and $\boldsymbol{B}(t,f)$, that is, $\boldsymbol{G}(t,f) = \boldsymbol{A}^{-1}(t,f)$ and $\boldsymbol{K}(t,f) = \boldsymbol{B}^{-1}(t,f)$.

Assuming that $X(t,f)$ and $Z(t,f)$ from (25) can be identical to that from (26) [35], equations (25) and (26) are combined to yield

$$\begin{pmatrix} E_1(t,f) \\ Y(t,f) \\ E_2(t,f) \end{pmatrix} = \begin{pmatrix} G_{xx}(t,f) & 0 & G_{xz}(t,f) \\ 0 & 1 & 0 \\ G_{zx}(t,f) & 0 & G_{zz}(t,f) \end{pmatrix}^{-1}$$

$$\times \begin{pmatrix} K_{xx}(t,f) & K_{xy}(t,f) & K_{xz}(t,f) \\ K_{yx}(t,f) & K_{yy}(t,f) & K_{yz}(t,f) \\ K_{zx}(t,f) & K_{zy}(t,f) & K_{zz}(t,f) \end{pmatrix} \begin{pmatrix} E_3(t,f) \\ E_4(t,f) \\ E_5(t,f) \end{pmatrix} \quad (27)$$

$$= \underbrace{\begin{pmatrix} \Re_{xx}(t,f) & \Re_{xy}(t,f) & \Re_{xz}(t,f) \\ \Re_{yx}(t,f) & \Re_{yy}(t,f) & \Re_{yz}(t,f) \\ \Re_{zx}(t,f) & \Re_{zy}(t,f) & \Re_{zz}(t,f) \end{pmatrix}}_{\boldsymbol{\Re}(t,f)} \begin{pmatrix} E_3(t,f) \\ E_4(t,f) \\ E_5(t,f) \end{pmatrix}$$

where $\boldsymbol{\Re}(t,f) = \boldsymbol{G}^{-1}(t,f)\boldsymbol{K}(t,f)$. The time-dependent spectrum of $E_1$ can thus be decomposed into the following three parts based on (27)

$$S_{E_1}(t,f) = \Re_{xx}(t,f)\tilde{\Sigma}_{xx}(t)\Re_{xx}^*(t,f) + \Re_{xy}(t,f)\tilde{\Sigma}_{yy}(t)\Re_{xy}^*(t,f)$$
$$+ \Re_{xz}(t,f)\tilde{\Sigma}_{zz}(t)\Re_{xz}^*(t,f)$$
$$(28)$$

where the upper script '*' denotes complex conjugate and transpose of a matrix, the first term can be regarded as the intrinsic power and the remaining two terms represent the combined causal relations from $Y$ and $\varepsilon_2$. Hence the causality from $Y\varepsilon_2$ to $\varepsilon_1$, namely the final expression for time-varying spectral GC $GC_{Y\to X|Z}(t,f)$ becomes

$$GC_{Y\to X|Z}(t,f) = GC_{Y\varepsilon_2\to\varepsilon_1}(t,f) = \ln\frac{|S_{E_1}(t,f)|}{|\Re_{xx}(t,f)\tilde{\Sigma}_{xx}(t)\Re_{xx}^*(t,f)|} \quad (29)$$

Note that the spectral function in (29) is a continuous function of frequency $f$, and can be applied to measure the spectral causality at any desired frequency from 0 up to the Nyquist frequency $f_s/2$. Generally the frequency resolution is not infinite, but relevant to the associated parameter approximations and underlying model order. A hypothesis test is required to determine whether the causal interaction in the stochastic processes is significant. The thresholds for statistical significance are computed from surrogate data by a permutation procedure under a null hypothesis of no interdependence at the significance level $p < 10^{-6}$.

The new proposed method for TF-CGC decomposition can now be summarized as bellow:

1) Set up the multivariate TVARX models (1) and (2), which are to be identified for TF-CGC analysis; expand all the time-varying parameters in each model using multiple B-spline basis functions and construct the corresponding time-invariant regression models (6).

2) Calculate $\{\bar{x}^v\}_{v=1}^d$ and $\{(\bar{u}_{n,i,k}^r)^v\}_{v=1}^d$ by modulating the output signals and expanded terms with the normalized test functions $\{\bar{\omega}^{(v)}\}_{v=1}^d$ and then get the ULS problem (11).

3) Select the significant term with the largest $RERR^0$ value as the first term and remove the selected expanded terms from the candidate dictionary; repeat the process and choose the $\varsigma$-th term by orthogonalizing all remained expanded terms with the $\varsigma - 1$ selected terms and calculating the associated $RERR^0$ value, and the term with the largest value is selected.

4) Determine the number of model terms using the APRESS statistic given in (17).

5) Approximate the coefficients of the selected model terms, and estimate the initial time-varying parameters using formula (5), hence the essential TVARX models for TF-CGC decomposition can now be established.

6) Normalize the bivariate and trivariate TVARX models by $P(t)$ and $Q(t)$ respectively to make the noise variables independent with each other, and calculate the spectrum representation of these normalized models.

7) Achieve the calculation of TF-CGC according to (28) and (29), and the statistical GC threshold is also estimated to get the significant TF-CGC relations.

### III. SIMULATIONS AND EXPERIMENTS

In this section, the performance of the proposed UROLS-based TF-CGC approach is firstly demonstrated using two simulation examples with various aspects of feature dimensions relative to cortical activities, and the effectiveness is compared with other three parametric methods: RLS, OLS and ROLS.

The proposed method is then applied to real EEG signals at scalp- and source-level. Specifically, *alpha* band (8~14 Hz) transient causalities among five-electrode scalp EEG signals recorded during MI tasks are analyzed. Furthermore the corresponding EEG source waveforms reconstructed at five MI related cortical regions of interest (ROIs) are also studied to detect 8~30 Hz dynamic causal activities in the neocortical sensorimotor network.

*A. Simulations and results*

*1) Time-frequency GC detection*

Consider the following TVARX processes

$$x(t) = 0.59x(t-1) - 0.2x(t-2) + a_1(t)y(t-1) + a_2(t)z(t-1) + e_x(t)$$
$$y(t) = 1.58y(t-1) - 0.96y(t-2) + e_y(t) \quad (30)$$
$$z(t) = 0.60z(t-1) - 0.91z(t-2) + e_z(t)$$

where $e_x(t)$, $e_y(t)$, $e_z(t)$ are three independent white noises with zero means and variances $\sigma^2(e_x(t)) = \sigma^2(e_y(t)) = 0.01$, $\sigma^2(e_z(t)) = 0.001$. The discrete time index $t$ is set to be equivalent to a sampling rate of 200 Hz, and each process consists of 2000 data points (i.e. $f_s = 200Hz, 0 \le t \le 2000, 0 \le t/f_s \le 10s$). $a_1(t)$ and $a_2(t)$ are time-varying strengths of interactions shown in Fig. 1 (a). In this example, the process $x(t)$ is influenced by $y(t)$ through $a_1(t)$ with fast oscillating strength, and is also interacted by $z(t)$ with continuously increasing intensity in the first half of the process and decreasing intensity in the second half.

The 3~6-th order B-splines with scale index $j = 4$ (i.e. $\xi_{k,j}^r : r = 3,4,5,6; j = 4$) are used to estimate the oscillating and continuous varying parameters of the model. The output signal and all the expanded terms are modulated with the first and second order derivatives of the cubic B-splines as test functions. Then the UROLS algorithm aided by APRESS is applied to construct the parsimonious model structure and recover the associated parameters. Based upon the identified TVARX models, time-varying causal influences from $y(t)$ and $z(t)$ to $x(t)$ in TF domain are calculated by means of the proposed parametric TF-CGC method. The detected TF-CGC distributions are given in Fig. 1(f). For comparison, the models in (30) are also estimated using the following algorithms: the standard RLS algorithm (with forgetting factor 0.94), the conventional OLS and ROLS algorithms with B-splines; the corresponding TF-CGC detection results are shown in Fig. 1(c)-(e), respectively. Fig. 1(b) represents the theoretical values of TF-CGC.

Fig. 1(c) shows that the traditional RLS method reflects monotonous changing interaction from $z(t)$ to $x(t)$ but fails to track oscillatory varying connectivity from $y(t)$ to $x(t)$. Fig. 1(d) indicates that the parametric TF-CGC measure using OLS with B-splines can detect the oscillating as well as ramp-shaped variations in causal influences but also produces false positive values at the wrong frequency without a desirable TF precision. Fig. 1(e) gives the causality obtained from ROLS with B-splines. The designed two types of varying influences are reflected in the results with almost no false positive values, but the causalities at some TF points are not detected and the measurements are much smaller than the theoretical values. The results shown in Fig. 1(d) and (e) measured on the basis of OLS and ROLS can be explained as a result of over-fitting and under-fitting of signal models in GC detection, respectively. In contrast, the proposed TF-CGC method using UROLS with B-splines aided by APRESS (Fig. 1(f)), can better reveal the dynamic interactions containing both fast oscillating and smooth continuous causal variations at almost all time and frequency points with high temporal and spectral precision.

The effectiveness of the proposed method is further quantitatively verified by analyzing the value of the mean absolute error (MAE), root mean squared error (RMSE) and peak signal to noise ratio (PSNR) of the TF-CGC measurements with respect to the corresponding theoretical values defined as

$$MAE = \frac{1}{NF} \sum_{t=1}^{N} \sum_{f=1}^{F} \left| \hat{C}(t,f) - C(t,f) \right| \quad (31)$$

$$RMSE = \sqrt{\frac{1}{NF} \sum_{t=1}^{N} \sum_{f=1}^{F} \left\| \hat{C}(t,f) - C(t,f) \right\|^2} \quad (32)$$

$$PSNR = 20\log_{10}(MAX / RMSE) \quad (33)$$

where $\hat{C}(t,f)$ is the measurements of TF-CGC $C(t,f)$ at each time and frequency point, $N$ is the data length and $F$ is the frequency range, $MAX$ denotes the maximum strength of the corresponding theoretical GC distribution. The associated results are given in Table 1. It is obvious that the calculated MAE, RMSE values of the proposed method are smaller than other three methods and the corresponding PSNR values are the largest one among four approaches, which statistically validate that the proposed scheme possesses better ability for tracking dynamic connectivity in both temporal and spectral domain.

TABLE I
A COMPARISON OF THE DETECTION RESULTS FOR EXAMPLE (1)

| TF-CGC | Method | MAE | RMSE | PSNR |
|---|---|---|---|---|
| $GC_{Y \to X|Z}(t,f)$ | RLS | 0.3357 | 0.8212 | 20.6469 |
|  | OLS with B-splines | 0.2179 | 0.5705 | 23.8114 |
|  | ROLS with B-splines | 0.1816 | 0.4997 | 24.9615 |
|  | **UROLS with B-splines** | **0.1642** | **0.4269** | **26.3295** |
| $GC_{Z \to X|Y}(t,f)$ | RLS | 0.2057 | 0.4541 | 21.5172 |
|  | OLS with B-splines | 0.1936 | 0.4209 | 22.1777 |
|  | ROLS with B-splines | 0.1630 | 0.4016 | 22.5839 |
|  | **UROLS with B-splines** | **0.1578** | **0.3438** | **23.9337** |

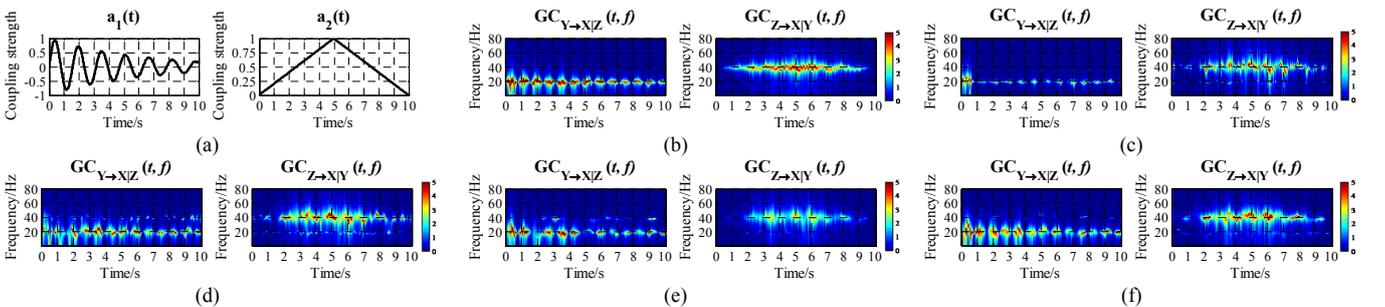

Fig. 1. (a) shows the time courses of dynamic coupling strengths in model (30), and (b)-(f) are the corresponding TF-CGC detection results employing different methods. Specifically, the theoretical values are given in (b); (c) the estimates using RLS with forgetting factor 0.94; (d) the estimates using OLS with B-splines; (e) the estimates using ROLS with B-splines; (f) the estimates using UROLS-APRESS with B-splines.

*2) TF-CGC detection under different conditions*

Consider a three-node network with nonstationary stochastic processes jointly described by the following TVARX models

$$x(t) = 0.53x(t-1) - 0.8x(t-2) + e_x(t)$$
$$y(t) = 0.53y(t-1) - 0.8y(t-2) + b_1(t)x(t-1) + e_y(t) \quad (34)$$
$$z(t) = 0.53z(t-1) - 0.8z(t-2) + b_2(t)y(t-1) + e_z(t)$$

where $e_x(t)$, $e_y(t)$ and $e_z(t)$ are Gaussian distributed noises with zero means and nonzero variances $\sigma_e^2$, $b_1(t)$ and $b_2(t)$ are time-varying coupling strengths from $x(t)$ to $y(t)$ and from $y(t)$ to $z(t)$ respectively, and the time index $t$ is assumed to be equivalent to $f_s = 200Hz$. Letting noise variances $\sigma_e^2 = 0.01$ and setting the coupling strengths vary according to the profiles given in Fig. 2(a), 20 trials of data with each trial containing 1000 points are produced in this case. Fig. 2(a) also illustrates the diagram of connectivity among the simulated three nodes.

According to Fig. 2(a), $x(t)$ has a causal relation on $y(t)$ in the first half of the simulation time interval, and $y(t)$ drives $z(t)$ in the second half in turn, moreover the dashed arrow means that $x(t)$ has an indirect effect on $z(t)$ mediated by $y(t)$. The corresponding theoretical values of TF-CGC are given in Fig. 2(b). In the ideal case, except the piece-wise varying immediate impacts of $x(t)$ to $y(t)$ and $y(t)$ to $z(t)$ with nonzero values, the other influences should be zero. Fig. 2(c) shows the CGC analysis results in TF domain based on the RLS algorithm with forgetting factor 0.90. The TF-CGC results detected by the parametric methods using the OLS with B-splines, the ROLS with B-splines, and the proposed UROLS with B-splines aided by APRESS are displayed in Fig. 2(d)-(f), respectively.

The classical RLS method (Fig. 2(c)) generates incorrect reflections of the nonzero dynamic connectivity ($GC_{X\to Y|Z}(t,f)$ and $GC_{Y\to Z|X}(t,f)$) in addition to the spurious information leakages [36] among other signal pairs predicted to be zero ($GC_{X\to Z|Y}(t,f), GC_{Y\to X|Z}(t,f), GC_{Z\to X|Y}(t,f)$, and $GC_{Z\to Y|X}(t,f)$). The method is sensitive to noises and cannot overcome the effect of mutual sources due to the limited convergence speed. This problem is partly solved by the OLS algorithm (Fig. 2(d)), although spurious interactions ascribed to over-fitting and leakages caused by mutual sources still exists. The parametric ROLS detection (Fig. 2(e)) alleviates the leakage issue to a negligible level, but the connection strengths from $x(t)$ to $y(t)$ and from $y(t)$ to $z(t)$ are much smaller than the theoretical values due to under-fitting. By comparison, the proposed procedure (Fig. 2(f)) can correctly detect the indirect impacts with zero strengths and well reflect the piece-wise variations in causality. All this show that the proposed approach appears to provide the most desired presentation of the connectivity.

The MAE, RMSE and PSNR of the TF-CGC estimates are presented in Table 2. Obviously, the proposed scheme has a better measuring performance for both abruptly changing direct impacts and indirect influences compared with other three approaches, demonstrating the advantage of the proposed TF-CGC method using ROLS-APRESS with B-splines in causality analysis for the coupling nonstationary systems.

In order to test the robustness of the proposed method in detecting connectivity, the previous procedures of signal generation are repeated under the following conditions: {Noise intensity: $\sigma_e^2 = 0.01, 0.1, 1$; Trial number: 10, 20, 30, 40, 50, 60, 70, 80, 90, 100}. Note that the levels chosen for both noise intensity and trial number cover the typical range for the cortical activity estimated by EEG technique. The RMSE values of the TF-CGC estimates for nonzero impacts ($GC_{X\to Y|Z}(t,f)$ and $GC_{Y\to Z|X}(t,f)$) under the three noise cases with different trials are shown in Fig. 3. The proposed framework outperforms the other three methods in all conditions. Specifically, the OLS algorithm could still fit to noises under high noise intensity ($\sigma_e^2 = 1$) thus makes the time-varying models over-fitting and leads to erroneous interaction values; the ROLS might perform well when samples are highly noisy but produces under-fitting models when the noise variances are small ($\sigma_e^2 = 0.01$). The proposed UROLS-APRESS approach, however, can effectively identify the essential TVARX models regardless of noise and trial amount circumstances hence obtains the best performance for analyzing dynamic TF connectivity patterns.

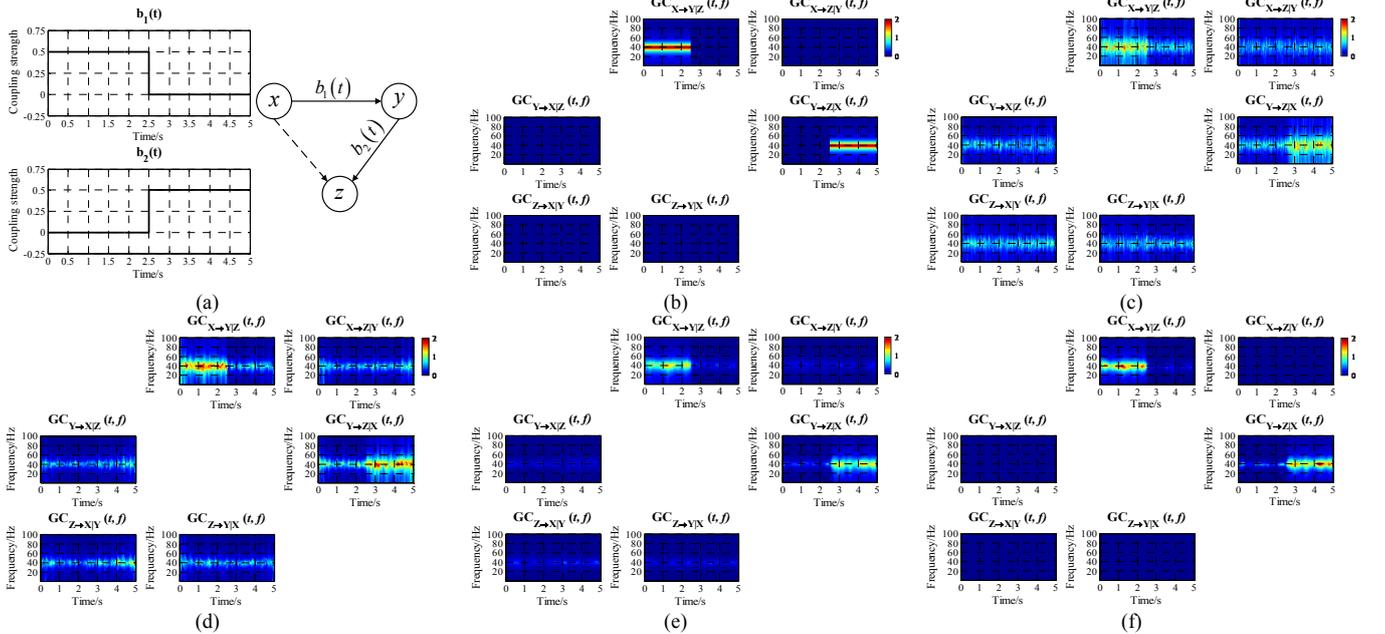

Fig. 2. (a) gives the time courses of dynamic coupling strengths and the diagram of interactions for example (2), and the corresponding TF-CGC detection results employing different methods ($\sigma_e^2 = 0.01$, 20 trials) are given in (b)-(f). Specifically, the theoretical values are shown in (b); (c) estimates using RLS with forgetting factor 0.90; (d) estimates using OLS with B-splines; (e) estimates using ROLS with B-splines; (f) estimates using UROLS-APRESS with B-splines.

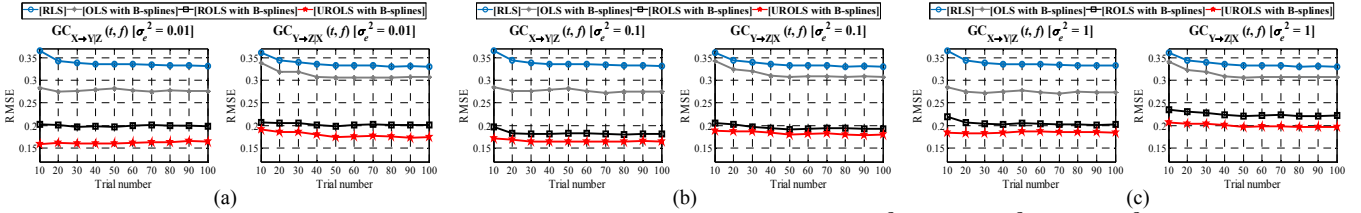

Fig. 3. The comparisons of dynamic TF-CGC estimates under three noise cases with different trials. (a) $\sigma_e^2 = 0.01$; (b) $\sigma_e^2 = 0.1$; (c) $\sigma_e^2 = 1$.

TABLE II
A COMPARISON OF THE DETECTION RESULTS FOR EXAMPLE (2)

| TF-CGC | Method | MAE | RMSE | PSNR | TF-CGC | Method | MAE | RMSE | PSNR |
|---|---|---|---|---|---|---|---|---|---|
| $GC_{X \to Y|Z}(t,f)$ | RLS | 0.2783 | 0.3431 | 15.3117 | $GC_{X \to Z|Y}(t,f)$ | RLS | 0.2511 | 0.2996 | 16.4902 |
| | OLS with B-splines | 0.2050 | 0.2744 | 17.2535 | | OLS with B-splines | 0.1917 | 0.2738 | 17.2714 |
| | ROLS with B-splines | 0.0933 | 0.2016 | 19.9310 | | ROLS with B-splines | 0.0437 | 0.0647 | 29.8067 |
| | **UROLS with B-splines** | **0.0743** | **0.1616** | **21.8529** | | **UROLS with B-splines** | **0** | **0** | **∞** |
| $GC_{Y \to X|Z}(t,f)$ | RLS | 0.2402 | 0.3007 | 16.4568 | $GC_{Y \to Z|X}(t,f)$ | RLS | 0.2744 | 0.3440 | 15.2885 |
| | OLS with B-splines | 0.1711 | 0.2600 | 17.7201 | | OLS with B-splines | 0.2289 | 0.3188 | 15.9510 |
| | ROLS with B-splines | 0.0399 | 0.0624 | 30.1166 | | ROLS with B-splines | 0.0969 | 0.2060 | 19.7442 |
| | **UROLS with B-splines** | **0** | **0** | **∞** | | **UROLS with B-splines** | **0.0858** | **0.1853** | **20.6628** |
| $GC_{Z \to X|Y}(t,f)$ | RLS | 0.2592 | 0.3264 | 15.7450 | $GC_{Z \to Y|X}(t,f)$ | RLS | 0.2330 | 0.2928 | 16.6892 |
| | OLS with B-splines | 0.1882 | 0.2881 | 16.8296 | | OLS with B-splines | 0.1688 | 0.2506 | 18.0422 |
| | ROLS with B-splines | 0.0513 | 0.0833 | 27.6080 | | ROLS with B-splines | 0.0386 | 0.0643 | 29.8595 |
| | **UROLS with B-splines** | **0** | **0** | **∞** | | **UROLS with B-splines** | **0** | **0** | **∞** |

*B. Applications to MI-EEG data at scalp- and source-level*

In this work, the Physiobank Motor/Mental Imagery (MMI) database [37] is used to evaluate the performance of the proposed TF-CGC approach. Specifically, the EEG dataset consists of 109 subjects performing different MI tasks while 64-channel signals were recorded based 10-10 systems, sampled at 160 Hz. The blocks where subjects imagined movements of left- and right-hand are selected in this study. Subjects performed a total of 45 trials and imagined one of the two tasks for a duration of 4 s in these chosen blocks. Three electrodes (T9, T10 and IZ) are discarded in the analysis, since they are spatial outliers relative to the other 61 electrodes which cover the scalp in an approximate uniformly distributed manner.

In order to estimate the EEG signals covering *alpha* rhythm, the noise-assisted multivariate empirical mode decomposition (NA-MEMD) algorithm is employed to decompose the 61-channel EEG data with two additional noise channels (SNR = 20dB, SNR = 40dB) [38]. The intrinsic mode functions (IMFs) prepared for the subsequent TF-CGC analysis are then determined based on the Hilbert-Huang spectrum of each obtained IMFs, where the ones most relevant to *alpha* rhythm are retained. For each trial, the mean of the pre-stimulus samples with duration of 2 s are subtracted for baseline correction, and the stimulus-triggered ensemble average is removed to mitigate the effect of inter-trial variations and the nonstationarity embodied in the mean [39].

*1) TF-CGC analysis of scalp EEG signals during MI tasks*

The proposed TF-CGC scheme is performed on scalp EEG signals to analyze the spectral specificity and temporal evolution of dynamic network interactions in *alpha* band during MI tasks. Five most commonly studied EEG channels in MI related researches (FZ, C3, CZ, C4, PZ) are used in this study. Setting these five channels as the network nodes for causality analysis, the net causal flows are then estimated by subtracting the calculated causal influences into the node from that out of the node: $CF_{node} = \sum_{i_c=1}^{N_n}(G_{node \to i_c} - G_{i_c \to node})$, where $N_n$ is the total number of nodes in a network and $G$ is the band integrated Granger causality, with self-causality assumed to be zero. The positive $CF$ denotes the net outgoing causal information flow away from the node (causal source), and the negative $CF$ refers to the net incoming flow towards the node (causal sink).

The trial signals of one participant are randomly selected from the dataset; setting the stimulus onset time as 0 s, Fig. 4(a) shows the significant net causal flow of C3 and C4 within 0~2 s during left-hand MI, represented as a function of time and frequency. The topographical maps of causal flow for the five-node network obtained by averaging $CF$ across temporal and spectral domain, with the time interval of 0.25 s and frequency range of 8~14 Hz (*alpha* band), are given in Fig. 4(b). The associated results for right-hand MI are represented in Fig. 5. For left-hand MI tasks, C3 mainly functions as a target (sink), which implies that C3 receives comparatively stronger interactions from other channels compared to outflow to them, whereas C4 functions predominantly as a causal source. The interaction relations shows no significant changes after around 0.5 s. Compared with left-hand MI conditions, the causal flow patterns of C3 and C4 are reversed during right-hand MI responses. These dominant information flows for C3 and C4 channel evaluated by the proposed TF-CGC method are consistent with the reported results in MI related studies [5, 40].

*2) TF-CGC analysis of EEG source signals during MI tasks*

In order to validate the efficiency of the new TF-CGC method in the context of a well-established interpretational framework, we further apply the method to EEG source signals during MI tasks to find the directed connectivity patterns in the neocortical sensorimotor network. Specifically, EEG-sources are firstly reconstructed based on ERPs, and the TF-CGC decomposition is further performed on the estimated single-trial source waveforms for source-level connectivity analysis.

The preprocessed 61-channel EEG data from each participant are averaged across trials to arrive at ERPs for left- and right-hand MI of 109 subjects. The 109 ERPs for two MI conditions are used in the exact low resolution electromagnetic tomography (eLORETA) to reconstruct EEG sources on the cortical surface [41]. The computations for inverse solution in eLORETA are implemented in a realistic head model based on the MNI152 (Montreal Neurological Institute) template, with the three-dimensional solution space restricted to cortical gray matter, as determined by the probabilistic Talairach (TAL) atlas. An entire 6239 cortical gray matter voxels with 5 mm spatial resolution constitute the solution space. EEG-source reconstruction at the whole brain level (all 6239 cortical voxels) is calculated, and a voxel by voxel comparison between left- and right-hand MI conditions in 8~30 Hz is also performed.

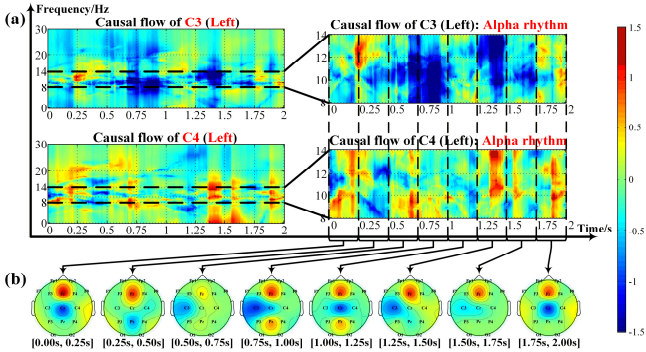

Fig. 4. TF-CGC analysis of scalp EEG signals during left-hand MI tasks: (a) net causal flows of C3 and C4 in TF domain; (b) topographical maps of causal flow for the five-node network consists of FZ, C3, CZ, C4 and PZ channels.

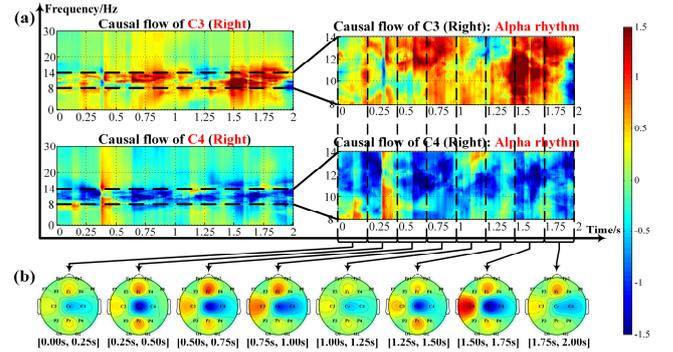

Fig. 5. TF-CGC analysis of scalp EEG signals during right-hand MI tasks: (a) net causal flows of C3 and C4 in TF domain; (b) topographical maps of causal flow for the five-node network consists of FZ, C3, CZ, C4 and PZ channels.

According to the results of statistical comparison, five ROIs are defined as network nodes for the next connectivity analysis (Table 3), where the single voxel nearest to the location of significant cortical activity (with extreme $p < 0.02$ in the $t$-test) is chosen to form each ROI. Time series of electric neuronal activity at these five ROIs are estimated with eLORETA. Based on the obtained single-trial source waveforms, the causality analysis in 8~30 Hz (covering both the *alpha* and *beta* rhythm) during 0~2 s, under left- and right-hand MI conditions, are then performed using the proposed TF-CGC method.

TABLE III
REGION OF INTERESTS (ROIS) USED FOR ANALYSIS OF TF-CGC

| ROIs | TAL coordinates | | | Anatomical regions | Brodmann areas (BAs) |
|---|---|---|---|---|---|
| | X | Y | Z | | |
| 1 | -5 | -13 | 28 | Limbic system | 23 |
| 2 | -50 | -17 | 42 | Primary somatosensory cortex (Left) | 3 |
| 3 | 50 | -17 | 42 | Primary somatosensory cortex (Right) | 3 |
| 4 | -45 | -18 | 38 | Primary somatosensory cortex (Left) | 3 |
| 5 | 45 | -18 | 38 | Primary somatosensory cortex (Right) | 3 |

The source signals from the same participant chosen in the scalp-level connectivity analysis are used for the evaluation at source-level, and the activities of five ROIs listed in Table 3 are marked as the nodes of causal network. For convenience, denote ROI 1 as Limbic Lobe or BA23, ROI 2 as Parietal Lobe.L1 or BA3.L1, ROI 3 as Parietal Lobe.R1 or BA3.R1, ROI 4 as Parietal Lobe.L2 or BA3.L2, and ROI 5 as Parietal Lobe.R2 or BA3.R2 in the following analysis. The results of TF-CGC and net causal flow associated with left-hand MI are shown in Fig. 6, where panel (a) presents the significant TF-CGCs, with dashed boxes indexed by 1 and 2 to outline the influences from the regions located in left hemisphere to those in right hemisphere and from right to left, respectively; panel (b) shows the net causal flows for the five node network; and panel (c) gives the CGCs averaged across 8~30 Hz plotted as functions of time for bidirectional influences between left and right regions. The corresponding results during right-hand MI are given in Fig. 7.

For left-hand MI tasks (8~30 Hz), Fig. 6(a) reviews the following observations: (i) the conditional causal influences from right regions to left (dashed box 2) are stronger than that from left to right (dashed box 1) especially after around 0.5 s; (ii) the enhancement of causal relations over the ipsilateral areas ($GC_{BA3.L1 \rightleftarrows BA3.L2}(t,f)$) and the blocking of interactions over the contralateral scalp ($GC_{BA3.R1 \rightleftarrows BA3.R2}(t,f)$) are detected along with the timeframe; and (iii) the causalities out of Limbic Lobe are more obvious than interactions input to it, and Limbic Lobe exerts greater causal influences on left regions (ipsilateral, i.e. BA3.L1 and BA3.L2) than right areas (contralateral, i.e. BA3.R1 and BA3.R2). In contrast, from Fig. 7(a), the causal patterns for 8~30 Hz activity in right-hand conditions show that: the causalities from left areas to right (dashed box 1) are more significant than that from right to left (dashed box 2) after approximately 0.25 s; the ipsilateral increase and contralateral decrease are also reflected in strong $GC_{BA3.R1 \rightleftarrows BA3.R2}(t,f)$ and small $GC_{BA3.L1 \rightleftarrows BA3.L2}(t,f)$; and Limbic Lobe outputs more evident causal influences to the ipsilateral sites (BA3.R1 and BA3.R2) than contralateral regions (BA3.L1 and BA3.L2). Fig. 6(b) shows that in left-hand MI, the ipsilateral regions (BA3.L1 and BA3.L2) function mostly as targets whereas the contralateral sites (BA3.R1 and BA3.R2) become dominant sources with no significant changes over time and frequency. For right-hand MI (Fig. 7(b)), BA3.R1 and BA3.R2 located at ipsilateral areas function predominantly as targets, and BA3.L1 and BA3.L2 in contralateral areas function as sources. Additionally, Limbic Lobe is the prominent source in both left- and right-hand MI conditions within the mainly entire timeframe and frequency range. From panel (c) in Figs. 6-7, the decrease of band averaged CGC from the ipsilateral regions to contralateral regions and the increase of CGC in the opposite directions are observed in both MI tasks, where the differences between the bidirectional influences enhanced apparently during 0.25~0.45 s in left-hand conditions, and the corresponding discrepancies occurred in 0~0.25 s for right-hand MI.

The network patterns for TF-CGC distributions between all node-pairs are also evaluated to better understand the dynamic organization of the connectivity network in 8~30 Hz. Specifically, the band mean TF-CGCs in 8~30 Hz are averaged along the timeframe with the interval of 0.25 s, and the obtained causality network groups changed through time are given in Fig. 8(a) and (b), for left- and right-hand MI tasks, respectively. Besides, the CGC network graphs in left- and right-hand MI conditions, estimated by averaging TF-CGCs over 0~2 s and 8~30 Hz, are shown in Fig. 8(c) and (d), respectively, where significant CGCs are indicated by lines with arrowheads whose thickness denotes the magnitude of causal influence, and only unidirectional CGC with greater value is presented for each node-pair though the interactions are bidirectional. In conditions of left-hand MI, the patterns of CGC are significantly observed initially during 0.5~0.75 s, however, for right hand the CGCs are firstly shown apparently within 0.25~0.5 s. It is obvious that the causality distributions presented in network patterns are consistent with those shown in Figs. 6-7, and can be interpreted in terms of the known anatomical pathways linking these related areas under MI conditions [42, 43].

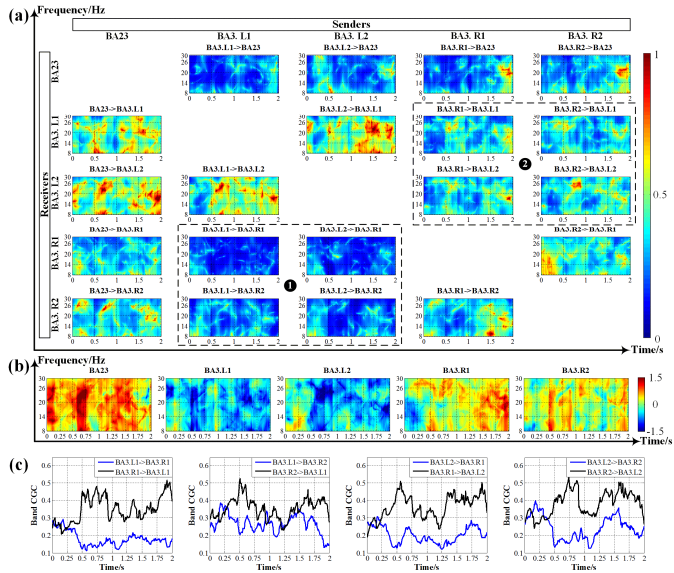

Fig. 6. TF-CGC analysis of EEG source signals during left-hand MI: (a) significant TF-CGC results among 5 ROIs; (b) associated net causal flows; (c) CGCs averaged in 8~30 Hz as functions of time for bidirectional interactions between left and right areas.

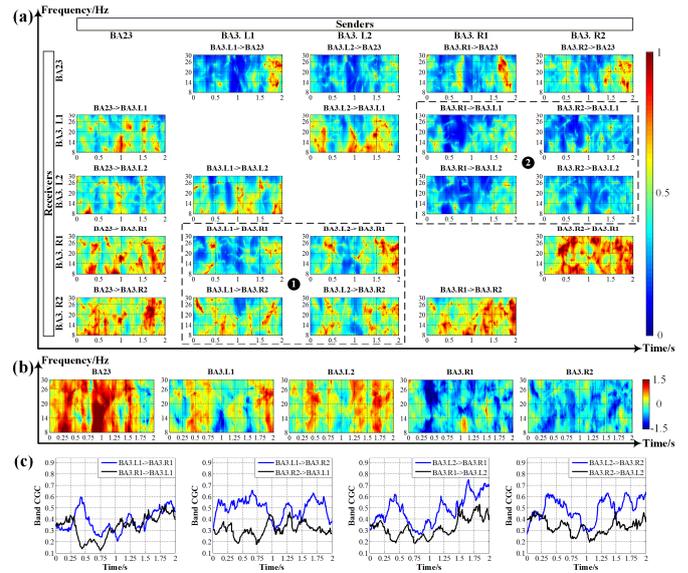

Fig. 7. TF-CGC analysis of EEG source signals during right-hand MI: (a) significant TF-CGC results among 5 ROIs; (b) associated net causal flows; (c) CGCs averaged in 8~30 Hz as functions of time for bidirectional interactions between left and right areas.

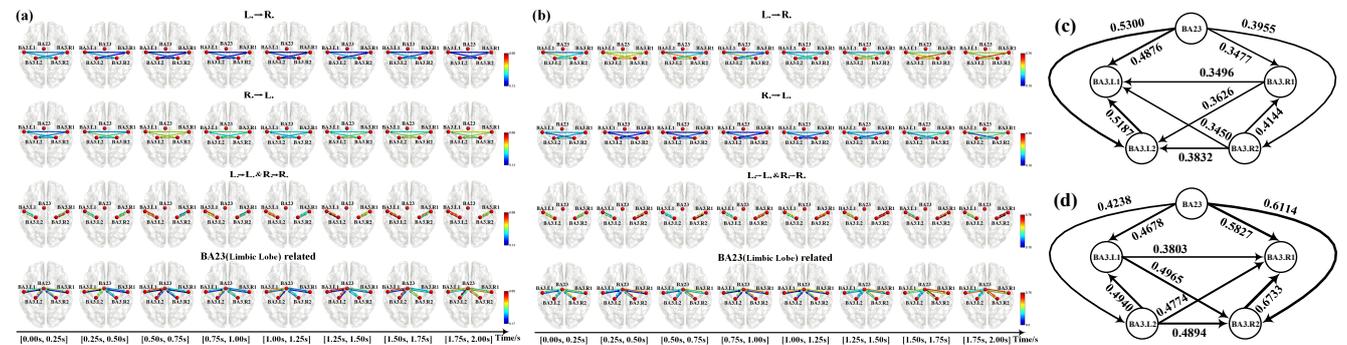

Fig. 8. Network patterns summarized TF-CGCs between all node-pairs: (a) causality network groups along time axis for left-hand MI; (b) causality network groups along time axis for right-hand MI; (c) CGC network graph for left-hand MI; (d) CGC network graph for right-hand MI.

## IV. DISCUSSION

Following the TF-CGC analysis of five scalp EEG signals (FZ, C3, CZ, C4, PZ), the obtained oscillatory causal networks in *alpha* band (Figs. 4-5) show that, for left- and right-hand MI tasks, the electrode signals recorded from opposite position function mainly as the causal source while the ipsilateral channel signals operate mostly as the net target almost within the whole 2 s after stimulus. These *alpha* oscillatory activities at scalp-level are in agreement with the direction of information flow postulated in the MI literatures [5, 40].

For connectivity analysis of MI-EEG source signals, the 8~30 Hz source-level network contains four active nodes symmetrically distributed on the left and right sides of the primary somatosensory cortex (BA3.L1, BA3.L2, BA3.R1 and BA3.R2) and a limbic area node (BA 23) (Table 3). Limbic system influences on motor behavior are widespread, and could range from the beginning of action to the motivational pace of motor output. More specifically, the area 24 and area 23 are the only part of the limbic lobe known to be interconnected in an organized fashion with motor related parts of the cerebral cortex [44]. Thus the inclusion of limbic node in the causal network is reasonable and necessary for this MI study.

Under both left- and right-hand MI conditions, three main conclusions observed from the *alpha* and *beta* band source CGC net (Figs. 6-8) are as follows. (i) First, considering the interhemispheric interactions, information flows from the contralateral sites to the ipsilateral regions are stronger than those in the reversed directions. (ii) Second, as for causal influences in the same hemisphere, an increase of information flows between ipsilateral regions and a decrease of those between contralateral areas are obviously detected. (iii) Third, for connectivity between limbic lobe and motor cortices, the limbic node functions as a causal source and exerts relatively greater causal influences on the ipsilateral sites compared with the contralateral regions. The first result of causal influence patterns is consistent with the well-known lateralization in hand MI tasks, specifically refers that, right- hand MI events activate primarily left hemispheric areas, whereas neural activity is lateralized to the right hemisphere for left-hand MI [45]. Besides, our study extends these knowledge from the point of view of the causal source, suggesting that the contralateral areas have the dominant function during MI in both left- and right-hand tasks. The second result agrees with the fact that, MI behavior can change oscillatory activities of the cortex and result in event-related synchronization (ERS, i.e. amplitude enhancement) in ipsilateral areas and event-related desynchronization (ERD, i.e. amplitude suppression) in contralateral regions of *alpha* and *beta* rhythms [46]. The result further suggests that this ERS and ERD phenomena can also be detected in effective connectivity within cortices. Moreover, the third finding can be interpreted by the function of limbic lobe, whose output contains one important component concerning its influence on motor related cortices [42]. Our results extend these findings through the causal network analysis, implying that limbic system impacts more on the regions ipsilateral than contralateral to the performing hand regardless of the left- and right-hand MI tasks.

In addition, note that the timing information for oscillatory network generated by the proposed method has a high resolution with temporal precision of 6.25 ms (i.e. the sampling interval), which cannot be reached by traditional sliding window approaches. Based upon the high TF resolution, the precise temporal and spectral patterns of source-level causal networks are obtained, and the latency of occurrence of MI tasks can thus be exactly examined. In this study, the onset time of stimulus evoked MI activation is determined as the time at which band averaged CGC in direction of left to right first began to deviate significantly from that in right to left. The results show that the onset times of left-hand MI tasks (0.25~0.45 s) slightly lag those under right-hand conditions (0~0.25 s), suggesting the influences of the asymmetry of the right-handedness on the directed connectivity networks during MI.

## V. CONCLUSION

A new parametric TF-CGC method is proposed for multivariate time-varying connectivity analysis in TF domain, where the UROLS-APRESS with multiwavelets is employed in generalized spectral CGC measure to achieve a high-resolution causality detection. Analyses on the simulation data show that the proposed approach can well detect both rapidly varying direct causalities and indirect effects among coupling systems over time and frequency. For real scalp and source MI-EEG data, the obtained connectivity patterns are physiologically and anatomically interpretable, and yield important insights into the dynamical organization of *alpha* and *beta* band cortical activities. An obvious advantage of the proposed method lies in its ability to track fast changing causal influences and eliminate indirect effects caused by mutual sources; these are attributed to the use of UROLS-APRESS. The novel application of the TF-CGC analysis to EEG signals can provide quantified and more detailed information of the underlying dynamic activities in oscillatory networks. Thus the procedure that oscillating networks coordinate activity between neocortical regions mediating sensory processing to arrive at motor perceptual decisions can be better understood through this study.